\documentclass[11pt]{JHEP}
\usepackage{amsmath, amsfonts,euscript,bbm}
\usepackage{epsfig, cite}
\usepackage{ifthen}
\usepackage{feynmf}
\usepackage{graphicx}
\usepackage{psfrag}

\newcommand{\Ket}[1]{\left|#1\right\rangle}
\newcommand{\KKet}[1]{\left|\left.#1\right\rangle\!\right\rangle}
\newcommand{\Bra}[1]{\left\langle#1\right|}
\newcommand{\BraKet}[2]{
  \left.\left\langle #1 \right| \! #2 \right\rangle
}
\newcommand{\Expect}[1]{\left\langle #1 \right\rangle}
\newcommand{\NOrder}[1]{\text{:}#1\text{:}}

\newcommand{\TF}[3]{\Theta_{#1}\bigl(#2\bigl|#3\bigr)}

\newcommand{\A}[1]{\mathcal{A}_{\rm #1}}

\def\tl{\tilde\lambda}
\def\tg{\tilde g}
\def\hf{\frac12}
\def\ap{{\alpha^{\prime}}}

\def\N{\mathcal N}
\def\Z{\mathbbm{Z}}

\def\DDD{\text{DD}\bar{\text{D}}}
\def\DD{\text{D}\bar{\text{D}}}

\def\at{\alpha^0}
\def\tat{\tilde\alpha^0}
\def\a{\alpha^0}
\def\ta{\tilde\alpha^0}

\def\N{\mathcal{N}}
\def\bk{{\mathbf k}}

\def\l{\lambda}

\title{On the Production of Open Strings from Brane Anti-Brane Annihilation}
\author{Louis Leblond\\
  Email: \email{lleblond@mail.lns.cornell.edu}
}

\abstract{We investigate the leading contribution to open string production in the time dependent background of the Brane Anti-Brane.  This is a 1-loop diagram and we use Boundary Conformal Field 
Theory (BCFT) techniques to study it. We show that the amplitude to a single open string na\"{\i}vely diverges when one looks at it as an expansion in oscillator levels. Nevertheless, we show that once we sum over all oscillator levels we get a finite result. We also clarify where to perform the inverse Wick rotation in this kind of problems.  This calculation could have important consequences for the theory of reheating in brane inflationary models.}  

\preprint{}

\begin{document}

\section{Introduction and Phenomenological Motivation}

Brane inflation is a powerful new paradigm in cosmology that might lead to 
interesting predictions
 \cite{Dvali:1998pa, Alexander:2001ks, Burgess:2001fx, Jones:2002cv, Garcia-Bellido:2001ky, Dvali:2001fw}.
If these predictions are observed experimentally they might
open windows into string theory itself.  We can put these predictions into two different types. 
First from the inflation potential, one can predict all the usual inflationary parameters measured
in the cosmic microwave background.  Althought interesting, it turns out that realistic models of brane inflation have enough freedom
and parameters to accommodate (without much fine-tuning in some cases) various power spectrum 
that fits the data \cite{sarah, Firouzjahi:2003zy, Firouzjahi:2005dh,Silverstein:2003hf, Alishahiha:2004eh, Chen:2004gc, Chen:2005ad, Chen:2005fe}. 
 Of course a plethora of different models of inflation fits the data
and so we must see these measurements not as an experimental test of string theory but rather as
constraints as to where in the landscape our universe inflated.  

Brane inflation also predicts how the universe should reheat. In the simplest models, inflation occurs when branes collide and annihilate.  These features of brane inflation cannot be described by usual quantum field theory and we can therefore hope to have purely stringy phenomena coming out of it.  The most famous prediction
of this type is the production of cosmic strings \cite{Sarangi:2002yt, Jones:2003da, Dvali:2003zh, Dvali:2003zj, Copeland:2003bj, Leblond:2004uc}
 at the end of brane inflation.  The 
production of cosmic strings can be shown to happen in $\DD$ annihilation from BCFT methods
\cite{Sen:2002nu}
or Boundary String Field Theory methods \cite{Kraus:2000nj, Takayanagi:2000rz,  Jones:2002si}.  Their properties could
shed interesting light into string theory \cite{Jackson:2004zg, Shlaer:2005ry}.  The phase transition at the end of 
brane inflation will not only produce defects but also particles which lead to reheating.  There has already been some study of reheating in the context of tachyon condensation for the brane anti-brane system 
\cite{Cline:2002it, Shiu:2002xp, Brodie:2003qv}. And recently, a more thorough analysis of reheating in KKLMMT-type models \cite{Kachru:2003sx} was performed in \cite{Kofman:2005yz, Frey:2005jk, Chialva:2005zy}. Additionally, a discussion of reheating in heterotic string theory models of inflation can be found in \cite{Becker:2005sg}.

The starting point of most of these calculations is that $\DD$ 
annihilates completely to non-relativistic massive closed strings \cite{Lambert:2003zr, Sarangi:2003sg}. 
Then they subsequently decay to standard model radiation.
Although this is certainly a legitimate starting point when there are no other branes around, it is not clear that it is correct when the brane annihilates with a stack of anti-branes. 
Indeed, one
might expect that such a collision should excite the spectator branes as well as producing 
closed strings.

Motivated by these phenomenological implications
 we would like to calculate the rate of production of open strings on a
spectator D-brane in the time-dependent background of a decaying
brane anti-brane ($\DD$) system.  The result should be compared to the
rate to massive closed strings already calculated
by Liu, Lambert and Maldacena (LLM)\cite{Lambert:2003zr}. 
In the LLM paper 
they found that the closed strings emitted are highly massive and
localized to the plane of the $\DD$ system.  The amplitude therein
goes like $\tfrac1{g_s} \times g_s = g_s^0$, the first factor from the
disc, the second from the closed string vertex operator.  The relevant
calculation uses the boundary state calculations of \cite{Sen:2002nu,
Larsen:2002wc} with an interacting BCFT and the insertion of a single
closed string vertex operator. 
\begin{figure}[h]
\centering
\includegraphics[width=3cm]{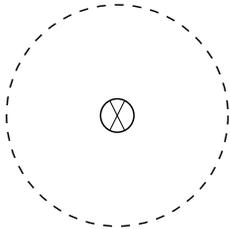}
\caption{Disc diagram with insertion of a closed string vertex operator (this is of order $g_s^0$).}
\end{figure}

If another brane is added to the decaying system, the time-dependent
background is expected to generate open string excitations.  This
would correspond to the first step of the reheating of the universe in $\DD$ models of
inflation, so it is important to estimate the rate and compare to the
rate to closed strings.

The leading open string creating process is a one-loop effect. Therefore,
 in this paper, we will extend the work of LLM to 1-loop. 
To see that open string creation is a 1-loop effect, 
note that the $\DDD$ system has states in
the Chan-Paton matrix (see for example \cite{Jones:2003ae}):
\begin{align*}
  \left(\begin{array}{cc}
    \left(\begin{array}{cc}
      A^{\bar{\text{D}}}& \bar T\\
      T& A^{\text{D}}
    \end{array}\right)& S^\dagger\\
    S& A
  \end{array}\right)
\end{align*}
Note that at lowest level, $S$ contains a $W$-type vector boson, and a
tachyon, but this will be unimportant; we take $S$ to indicate
\emph{any open string states} stretching between the decaying $\DD$
system and the spectator brane. Similarly, $A$ at lowest level would denote
the gauge field on the remaining D-brane but here we take it to be 
\emph{any open string states} living on the spectator brane.
The states $S$ cannot be in any final
state since they have one end on the decaying system, so only $A$ can
be a final state.  However, in order to produce $A$, $S$ must be
excited since there is no direct coupling between $A$ and $T$ which is
the time-dependent source.  Note that as for the decay to closed
strings, the $A^{\text{D},\bar{\text{D}}}$ strings and their
excitations are irrelevant and can be set to 0.  The lowest order
process to excite $A$ then is: 
\begin{figure}[h]
\centering
\psfrag{A}{$A$}
\psfrag{Sd}{$S^\dagger_i$}
\psfrag{S}{$S_i$}
\includegraphics[width=7cm]{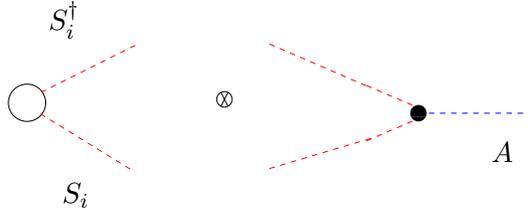}
\caption{$\DD$ (large white circle) emitting the fields $S$ and $S^\dagger$ that then annihilates on a 
spectator brane to produce a gauge field $A$.}
\end{figure}

As this is a 1 point function, on the spectator branes, the field A must
be a gauge singlet.
The corresponding string diagrams are disc amplitudes. When
summed over all possible intermediate states, $S_i$, they lead to the
annulus diagram with one boundary corresponding to the decaying $\DD$
system and the other boundary representing the spectator D-brane, with a
vertex operator insertion corresponding to $A$. The string
calculation is 

\begin{align}\label{amp}
\A{}(E) & = \Bra{B}\frac{1}{L_0 +\tilde L_0}\mathcal{V} \Ket{N},
\end{align}
where $\Bra{B}$ is the boundary state of the brane anti-brane system, 
$\mathcal{V}$ represents the vertex operator of the state produced and $\Ket{N}$ is
the usual Neumann boundary state for the spectator brane.

\begin{figure}[h]
\centering
\includegraphics[width=7cm]{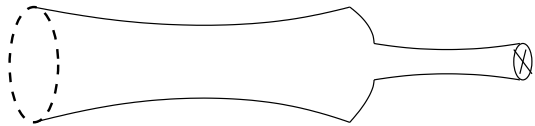}
\caption{1-point function on the cylinder where one end (dashed line) represents
the $\DD$ time dependent source and the other end represents the spectator brane.}
\end{figure}

This process then goes like $g_s^0 \times (g_s^{1/2}) = g_s^{1/2}$, where
the first factor comes from the annulus, and the second from the
open string vertex operator.  That this is higher in string coupling
na\"{\i}vely implies that this contribution should be subleading compared to closed string production.
On the other hand if one looks at pair production (which is now of order $g_s$), we are no
longer limited to gauge singlet and we will therefore have an enhancement from the number
of spectators branes.
\begin{figure}[h]
\centering
\includegraphics[width=7cm]{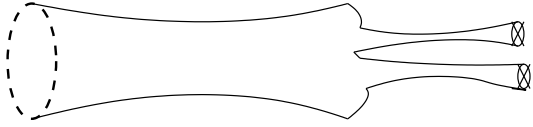}
\caption{2-point function on the cylinder where one end (dashed line) represents
the $\DD$ time dependent source and the other end represents the spectator brane.}
\end{figure}
Indeed, the square of the amplitude for the 2-point function goes like $g_s^2$ but when we sum over all
states we get equal contribution for all the states produced in the Chan-Panton matrix.  For an 
$SU(N)$ 
gauge theory we would get $N^2-1$ such states and therefore the controlling parameter is 
roughly $N^2g_s^2$.  It is then possible for sufficiently large N 
to imagine that the 2-point function would lead to a non
negligible production of open strings. 

In this paper we will concentrate on the one-point function as it is much simpler and exhibits the main properties and problems that one encounters also for the 2-point function. Nevertheless we will comment on the 2-point function later on as this is surely the most important diagram for open string production.  As we will discuss later, there is a diagram at 1-loop that will contribute to closed string emission and it will also be enhanced by the same factor of N.  Therefore, enhancement by factors of N is not necessarily an advantage of open string production versus closed strings but it does 
show that 
$g_s$ is not necessarily the controlling parameter. \footnote{We thank Xingang Chen for pointing this to us.}

These are the motivations behind these 1-loop calculations but this paper will mostly address the computational techniques needed and the problems associated with them. 
Indeed, it appears that the 1-loop amplitude in the time dependent background of $\DD$ is not well defined if one expands the amplitude in oscillator levels as we get terms that diverge at late time 
(as was noted in \cite{Chen:2003xq}).  Historically these terms were thought to cause problems for the tree-level calculation \cite{Okuda:2002yd} but \cite{Lambert:2003zr} showed that an appropriate choice of gauge leads to a finite tree-level amplitude.  In the rest of this paper we will show that these divergent terms do appear in the 1-loop amplitude and they cannot be gauged away.  Nevertheless we argue that the amplitude is finite and that all the divergent terms come in such a way that they sum to a finite result.  
We argue this by truncating Sen's boundary state to an infinite subset of terms and doing so we get
the following behavior for the 1-loop amplitude:

\begin{align*}
&\A{1\rm{pt}}(E) \approx \frac{1}{\sinh{\pi E}} w(E), 
& \A{2\rm{pts}}(E)\approx \frac{1}{\sinh{2\pi E}} w^\prime(E),
\end{align*}
where $w(E)$ and $w\prime(E)$ are unknown polynomials. 
This form for the amplitude would imply that open strings production is by itself non negligible but in general phase space consideration would still favor closed string emission over open string.  We cannot exactly predict how much open strings versus closed strings will be produced without a specific knowledge of
these polynomials but it appears that, due to phase space consideration (more phase space for closed strings), closed strings production will dominate 
in most cases.
   
In this paper we will first discuss more in depth how the problem with divergent terms arises, 
then we will look in detail at the 
calculation by first reviewing the BCFT techniques as well as the tree level calculation of \cite{Lambert:2003zr}.  We then look at a subset of terms contributing to the amplitude and show how these terms can be summed to something finite.  Finally we will conjecture about the final form of the amplitude and discuss its implications for the theory of reheating.

\section{Problematic Divergent Terms in Sen's Boundary State} 

The goal of this paper is to explore the properties of the
 1-loop amplitude in the time dependent background of $\DD$.
Before going into details for any calculations, let us first describe the main problems that we will encounter.

First of all there is a technical difficulty as the vertex operators of 
physical open string states are naturally expressed in the Virasoro basis 
($\alpha$) as opposed to the brane anti-brane boundary state
 which is naturally expressed in terms of the 
$SU(2)$ current algebra basis.
This clash of basis leaves two main routes of attack to perform the calculation.

The first line of attack comes from the realization that the boundary states only excite
open strings with integer values of momentum. This allows us to write the vertex operator either
as a product of $SU(2)$ current operators or as a product of fermions.  This way the full amplitude
can be evaluated for specific values of momenta and then we can sum over all momenta. 
This calculation turns
out to be well defined but technically intractable as the terms in the sum get quickly out
of control for large momentum.  We therefore merely present the basic idea in the appendix.

The second line of attack is to expand the boundary state in the $\alpha$ basis and calculate
the amplitude as a series in power of oscillator levels
\begin{align*}
\Ket{B} & = B_{0,0}\Ket{0} + B_{1,1} \a_{-1}\ta_{-1}\Ket{0}+ \cdots,\\
\A{} & = \A{0,0} +\A{1,1} + \cdots.
\end{align*}
This method is instructive as it shows the subtleties of the Wick rotation in such a system.  
We show that this series appears to be badly behaved
as individual terms at level 2 and beyond
seems to diverge at late time.
As we will see, choosing a special gauge is not going to solve the problem for the 1-loop calculation as it does for the tree level. 
On the other hand, by using a subset of terms in the boundary state expansion to all levels (namely 
all terms of the form $\a_{-n}\ta_{-n}$) we can show that part of 
these divergent pieces lead to a finite contribution once you sum 
over \emph{all momenta and all oscillator levels} before Wick rotating. 
 This essentially extends the original work of Sen and it is the main result of this paper. Although interesting, this does not tell us very much  about the amplitude as we have only chosen
a subset of all the strings running in the loop. Still, it does 
give tantalizing hints that the amplitude should be finite.

\section{Rolling Tachyon Boundary Conformal Field Theory}

The time-dependent BCFT representing a homogeneously evolving tachyon
on a D-brane in bosonic string theory was introduced by Sen
\cite{Sen:2002nu}. To be a solution to classical open string theory,
the insertion on the worldsheet boundary must be marginal, and
independent of oscillators to correspond to the tachyon.  One possible
deformation (considered by Sen) is therefore $\lambda \cosh{X^0/\sqrt{\ap}}$ 
(in the remainder of this paper, we shall work in units where $\ap = 1$)
\begin{align*}
S & = - \frac{1}{2\pi} \int d^2 z \partial_z X^0\partial_{\bar z} X^0 + \l \int d\tau \cosh{X^0}.
\end{align*}
By Wick rotating $X^0 \rightarrow iX^0$ , one
gets the 
celebrated Sine-Gordon action for which the boundary state is 
well known \cite{Callan:1994ub}. It turns out to be 
a simple $SU(2)$ rotation of the
Neumann boundary state
\begin{align*}
\Ket{B} &= e^{i\theta^a J^a}\Ket{N}\\
\Ket{B} & = \sum_j\sum_{m = -j}^j D^j_{m,-m}(R)\KKet{j,m,m}.
\end{align*}
where $R$ is a $SU(2)$ rotation matrix and $J^a$ are the $SU(2)$ current operator (see appendix).

This tachyon profile is often called the full S-brane and it represents 
finely tuned incoming radiation creating the $\DD$ system,
followed by its decay back into radiation. Of course the decay is the only
physically relevant part. The tachyon profile $\tl e^{X^0}$ ($\hf$ S-brane)
captures only the decay part and is therefore what we will use for 
the rest of this paper. 

The boundary state for the $\hf$ S-brane is a simple generalization of 
the full S-brane case where the rotation matrix is now an element of 
$SL(2,\mathcal{C})$. The explicit form is (with $\tl = 2\pi\lambda$)
 \cite{Gaberdiel:2004na}:
\begin{align}\label{Bdefinition}
  \Ket{B} \equiv \sum_{j\in \hf \Z^+}\sum_{m\ge 0}^j
  {j+m \choose 2m}(-\tl)^{2m}\KKet{j,m,m}.
\end{align}

Sen calculated those parts of the corresponding boundary state which
are oscillator-free, and proportional to $\at_{-1}\tat_{-1}$, in the
euclidean theory,
\begin{align}\label{BDefinition}
  \Ket{B} = \int dE \left\{ f_{\rm{Eu}}(E)
  + g_{\rm{Eu}}(E)\at_{-1}\tat_{-1}
  + \ldots \right\}\Ket{E}.
\end{align}
In the Wick rotated theory, $f_{\rm Eu}$ and $g_{\rm Eu}$ are
calculated either in a perturbative expansion of the boundary
interaction or by using the $SU(2)$ structure of the system,
\begin{align*}
  f_{\rm Eu}(E) &= \BraKet{E}{B}_{\rm Eu} 
  = \sum_{n=0}^\infty (-\tl)^n 2\pi\delta(n-E),\\
  g_{\rm Eu}(E) &= \Bra{E}\at_1\tat_1\Ket{B}_{\rm Eu} 
  = \sum_{n=1}^\infty (-\tl)^n 2\pi\delta(n-E)
  = f_{\rm Eu}(E) - 2\pi\delta(E).
\end{align*}
An inverse Wick rotation of these quantities would be somewhat ill
defined. Sen argues that it is sensible only to inverse Wick rotate the
appropriate, physically meaningful quantities: $f_{\rm Eu}(\tau)$ and
$g_{\rm Eu}(\tau)$, the Fourier transform of $f$ and $g$, which have meaning
as components of the stress-energy tensor of the decaying brane system
\cite{Sen:2002nu, Gaberdiel:2004na}.  Consequently,
\begin{align*}
  f_{\rm Eu}(\tau) = \int\frac{dE}{2\pi} e^{i\tau E}
  f_{\rm Eu}(E) &= \frac1{1+\tl e^{i\tau}}\\
  \to f(t) &= \frac1{1+\tl e^t}.
\end{align*}
Similarly,
\begin{align*}
  \to g(t) = 2\pi\left[\frac1{1+\tl e^t} -1\right],
\end{align*}
where we have done an inverse Wick rotation ($\tau = -it$). 
This procedure defines well the stress-energy tensor and the boundary
state, because $f$ and $g$ have well-defined Fourier transforms,
\begin{align*}
  &f(E) = \frac{i\pi \tl^{-iE}}{\sinh(\pi E)},
  &g(E) = f(E) - 2\pi\delta(E).
\end{align*}
However, when considering other states that should appear in
(\ref{BDefinition}), this is no longer true; for instance,
\cite{Okuda:2002yd, Constable:2003rc} calculate the coefficients of
$\at_{-2}\tat_{-2}$ and $(\at_{-1})^2\tat_{-2}$ in $\Ket{B}$ to
contain terms like
\begin{align}\label{BadIntegral}
  \sim \int dt\; e^{t+iE t}.
\end{align}
In \cite{Okuda:2002yd}, this was interpreted as the coupling of the
decaying brane system to massive closed strings diverging at late
times.  However, \cite{Lambert:2003zr} calculate the rate of
generation of closed strings from a decaying brane, and showed that the
rate to an individual closed string does not diverge. 

The leading diagram that contributes to the production of closed strings
is the disk diagram (see figure 1).
The corresponding calculation in the euclidean theory is 
\begin{align}
\A{\text{disk}}(E) & = \Bra{0} \mathcal{V}_c\Ket{B}.
\end{align}
It was argued in \cite{Evans:1998wq, Hwang:1991an}, 
that one can choose a gauge for the \emph{physical}
states such that there are no time-like oscillators in the vertex operator besides the
momentum part,\footnote{Note that this gauge choice only applies to physical closed
and open string states and it does not affect the boundary states, we are grateful to Nicolas Jones for clarifying this point.}
\begin{align}
\mathcal{V}_c & = e^{iEX^0} \mathcal{V}_{\text{sp}}.
\end{align}
The spatial part of the amplitude then turns out to be simply a 
phase and the timelike part simply picks up the part of the boundary
that contains no oscillators.  Hence we find that the amplitude is
\begin{align}
\A{\text{disk}}(E) & = \Expect{E|B},\\
& = 2\pi\sum_n (-\tl)^n \delta(E-n) = f_{\text{Euc}}(E).\nonumber
\end{align}
This is the amplitude (up to a phase) in the euclidean theory. By the same procedure as before we get in the lorenztian theory:
\begin{align}
\A{\text{disk}}(E) & = \int dt \frac{e^{iEt}}{1+\tl e^t}\\
& = \pi i \frac{e^{-iE\log{\tl}}}{\sinh{\pi E}} \nonumber
\end{align}

From this, we see that these divergent
pieces do not show up in the tree-level calculation as one can choose a
gauge for the emitted vertex where there are no oscillators in the
timelike direction.  By doing so, the only part of the boundary state
contributing to the amplitude is the zero mode part $f(E)$. 
Loosely speaking, one could say that 
these divergent terms can be 'gauged away' in the tree
level calculation. This also means that $f(E)$ is sufficient
to completely specify every physical amplitude of closed strings
at tree level. 

At one-loop one expects a different outcome. Indeed, the full
boundary state is needed as it is to be projected onto a
Neumann boundary state containing all levels of oscillators.
Choosing a special gauge for the emitted vertex will not prevent these
terms from appearing when we trace over the whole open string
spectrum. Nevertheless, it would be extremely surprising if the amplitude 
of the brane anti-brane system to decay to an \emph{individual} open
string was divergent. 

Another important issue that we encounter while going to a 1-loop analysis 
of tachyon condensation is that it is ambiguous where we should 
perform the inverse Wick rotation to go back to a Lorentzian signature.
As was pointed out recently \cite{deBoer:2003hd,Gaberdiel:2004na},
since terms like (\ref{BadIntegral}) are ill-defined, there must be
some {\it physical prescription to define them}.  After we inverse Wick rotate
to the Lorentzian frame, this could be given by the
condition of conformal invariance of the boundary state:
\begin{align}\label{BConformal}
  &(L_n - \tilde L_{-n})\Ket{B} = 0, &\forall n \in \Z.
\end{align}
Imposing this condition gives a way to define (\ref{BadIntegral}), via
\begin{align*}
  \Bra{E}(\at_1)^2\tat_2\Ket{B} 
  &= \frac1{E}\Bra{E} L_1 \at_1\tat_2\Ket{B} = \ldots\\
  \Rightarrow \left(1-\frac1{E\cdot E}\right)
  \Bra{E}(\at_1)^2\tat_2\Ket{B} &= 0,
\end{align*}
giving for $E\cdot E \ne 0,1$, $\Bra{E}(\at_1)^2\tat_2\Ket{B}=0$.  For
the {\it timelike system} we are considering, $E\cdot E = -E^2\le0$,
so as long as $E\ne0$, (\ref{BadIntegral}) can be thus defined to be
0.  As noted in \cite{deBoer:2003hd, Gaberdiel:2004na}, this example
reflects the fact that the $c=1$ timelike CFT has no null vectors (the
$c=1$ Kac-Moody determinant is nowhere vanishing for $h<0$).  
In this case it is
therefore sufficient to consider the boundary state of Ishibashi
states built over highest weight states of energy $E$, which captures
all the necessary information of the timelike system:
\begin{align}\label{BosonicB}
  &\Ket{B} = \int\limits_0^\infty dE\;
  f(E) \KKet{E},
  &\KKet{E} = \exp\left[-\sum_{n=1}^\infty 
    \frac1n \at_{-n}\tat_{-n}\right]\Ket{E}.
\end{align}
That is, the oscillator expansion of the boundary state is just that
of a free timelike boundary state with Neumann boundary conditions.
This reproduces $f$ and higher diagonal states correctly (only for $E
\ne 0$; terms at $E = 0$ are important in, for instance the
construction of the stress-energy tensor of the system
\cite{Sen:2002nu, Balasubramanian:2004fz}, and are omitted from
$\Ket{B}$ here), and sets the coefficients of all off-diagonal states
such as (\ref{BadIntegral}) to zero.  

This approach is correct for tree-level calculations and it is 
perhaps the clearest way 
of understanding why these divergent terms do not appear in the calculation
of \cite{Lambert:2003zr}.  On the other hand the boundary state (\ref{BosonicB}) 
\emph{should not} be used for 1-loop calculation.

Sen's prescription is to inverse Wick rotate only the physically meaningful 
quantities.  The boundary state is physically meaningful because when
projected on the closed strings vacuum it measures the 1-point 
tree level amplitude 
of closed strings from the decaying system.  However, this is a 
trick and we need to remember that the inverse Wick rotated
boundary state is only physically 
meaningful once it is projected on the vacuum.
For the 1-loop amplitude, we do not project the boundary state on
the vacuum and we should therefore not use the lorenztian boundary state 
(and its truncated but finite version (\ref{BosonicB})).
Indeed, the correct calculation is to \emph{stay until the very end in the euclidean
frame} and then inverse Wick rotate the final result.
Doing so, we will modify the sum leading to $f(x_0)$ and $g(x_0)$ and
we will get \emph{different} time dependence than what one gets at
tree level. 

This point is very important so let us elaborate more
on the issue.  It is quite clear that Sen's boundary state includes more than the simple finite diagonal 
terms \cite{Sen:2004zm}
\begin{align*}
\Ket{B} & = e^{\sum_{n>0} \frac{1}{n} \a_{-n}\at_{-n}}f(X^0)\Ket{0} +\Ket{\tilde B}.
\end{align*}
Now this boundary state is physically meaningful as it measures the 1-point function on the
disc so we can Wick rotate it.  As we do so we find ill-defined terms (\ref{BadIntegral}).  
One can see that these terms are not a problems in two different ways. First one can see that
in the Lorentzian theory, we can define these ill-defined terms to be 0 by using conformal
invariance.  Secondly, we can simply choose a gauge for the vertex operators such that there
are no oscillators in the time-like direction.  It is an interesting question as to whether the two 
methods are related. Both ways fail when we try to calculate 1-loop amplitudes. Indeed choosing
a special gauge (although probably useful for computational purposes) will not help as we
project on a Neumann boundary state which will pick up these terms anyway.  Furthermore, we cannot
use the Lorenztian boundary state as we need to do both the integral over moduli space 's' and
trace over all the closed strings exchanged between the branes. Each step only converges if we do
an analytic continuation to euclidean space forcing us to use the extra pieces of Sen's boundary 
state.   At this point, we have a legitimate question to answer: is the 1-loop amplitude finite?

As we will argue in the following section,  \emph{we need to use the euclidean boundary state and inverse Wick rotate the finite result}, 
yet the answer is finite as the integral over the moduli 's' together with summing over all levels conspire to sum
all these divergent terms to something finite.

\section{Expanding the 1-Loop Amplitude in Oscillator Levels.}

Now, as explained before, the ultimate goal is to evaluate the 1-loop amplitude in superstring
theory to any open string states. This is very hard and so we will first simplify to the simplest 
possible calculation.  We will work in the bosonic theory and calculate the amplitude only to the open
string tachyon.  Even then, we will further simplify by only considering part of the boundary state
$\Ket{B}$. We will first disregard the spatial part of the boundary state (and disregard the tachyon coming from it which is an artifact of bosonic string theory) as well as expand the boundary state 
in oscillator levels.  Doing so, 
our answer is not trustable to be a good approximation but the goal of this paper
is more conceptual then quantitative. We hope to convince the reader that the dangerous terms 
coming from the time-like part of the boundary state do not lead to infinities.

We will calculate the amplitude (\ref{amp}) in the euclidean theory for the simplest state where it becomes (we calculate
the complex conjugate for computational convenience):
\begin{align*}
\A{Eu}(E) & = \int dx_0 \Bra{N} e^{-iEX^0} \frac{1}{L_0 +\tilde L_0}\Ket{B},\\
\A{Eu}(E) & = \int ds \int dx_0 \Bra{N} e^{-iEX^0} q^{L_0}\Ket{B},
\end{align*}
where $q = e^{-2s}$ and we have used the conformal condition on $\Ket{B}$,
\begin{align}\label{BConformal}
  &(L_n - \tilde L_{-n})\Ket{B} = 0, &\forall n \in \Z.
\end{align}
Note that no analytic 
continuation is needed in the last step as $E\cdot E = E^2 > 0$ and the integral over 's' converges.
Now Sen's prescription tells us that we should inverse Wick rotate back only the fourier transform of 
$\A{Eu}(E)$
\begin{align}\label{theamplitude}
\A{Eu}(\tau) & = \int \frac{dE}{2\pi} \int dx_0 \int ds  e^{iE\tau} \Bra{N} e^{-iEX^0} q^{L_0}\Ket{B}.
\end{align}
The final answer in the lorenztian frame is obtained after we inverse Wick rotate $\tau = -it$ and inverse Fourier transform
\begin{align}
\A{Lor}(E) & = \int dt e^{iEt} \A{Lor}(t).
\end{align}
Note that the sign of the exponential is the same in both cases since we have done a Wick rotation.
As a first exercise, let us evaluate explicitly the one-loop amplitude as an expansion in terms of 
oscillator levels
\begin{align*}
  \Expect{\NOrder{e^{-iEX^0}}} =\int dx^0\int ds \Bra{N}\NOrder{e^{-iE X^0}} q^{p^2/4}
Q^{L_0^\text{osc}}\Ket{B}.
\end{align*}
Here, we have arbitrarily made
a distinction between oscillators and momentum ($Q=q=e^{-2s}$). The goal will be to evaluate
this amplitude as a series in Q.  There is no good reason to expect Q to be a 
good expansion parameter, it is just that the calculation is tractable this 
way. 

The boundary state for the rolling tachyon was worked out up to level (3,3) in 
\cite{Constable:2003rc} (and up to level (2,2) in \cite{Okuda:2002yd})
\begin{align*}
\Ket{B} = B^{(0;0}\Ket{0}+ B^{(1;1)}\a_{-1}\ta_{-1}\Ket{0}+\frac{1}{\sqrt{2}}B^{(1,1;2)}(\a_{-1})^2\ta_{-2}\Ket{0}+\cdots.
\end{align*}
We reproduce here table 4 of \cite{Constable:2003rc} where $\tg = -\tl e^{ix_0}$ and 
$f = \sum_{n=0}^{\infty} \tg^n$.

\begin{center}
\begin{tabular}{|c|c|}
\hline
$(\sigma ; \tilde\sigma)$ & $B^{(\sigma ;\tilde\sigma)}$\\
\hline
$(0;0)$ & $f$\\
\hline
$(1;1)$ & $f-2$\\
\hline
$(2;2)$ & $f-2 + \tg$\\
\hline
$(1^2;1^2)$ & $f  + \tg$\\
\hline
$(1^2;2)$ & $-\sqrt{2}\tg$\\
\hline
$(3;3)$ &  $ f-2 +\frac43\tg -\frac23\tg$\\
\hline
$(2,1;2,1)$ &$ f+\tg -2 \tg^2$\\
\hline
$(1^3;1^3)$ & $f -2 +\frac23\tg-\frac43\tg^2$\\
\hline
$(2,1;1^3)$ & $\sqrt{\frac23}[\tg + 2\tg^2]$\\
\hline
$(2,1;3)$ & $\frac{2}{\sqrt{3}}[-\tg+\tg^2]$\\
\hline
$(3;1^3)$ & $ \frac{2\sqrt{2}}{3}[-\tg -\tg^2]$\\
\hline
\end{tabular}\\
Table 1: Boundary state coefficients up to level (3,3)
\end{center}
The first thing to note in this table is that every possible left/right symmetric state
(equal level on both side) contributes in 
the boundary state contrary to the usual Neumann boundary state where only left/right identical states
contribute.  The function $f$ is common to most terms and leads to a 
finite contribution to the amplitude after we perform the sum in the euclidean theory. 
On the other hand, terms like $\tg$ will exponentially 
diverge at late time after Wick rotation.  As this can hardly be made sense of, it must be they that
somehow all conspire to sum to something finite. 

We will get hints that this will happen by using a closed formula found in \cite{Constable:2003rc} 
for all terms of the form $\a_{-N}\ta_{-N}$ 
\begin{align}\label{infinitesub}
B^{(N;N)} & = f -\frac{2}{N}\sum_{n=0}^{N-1} (N-n)\tg^n.
\end{align}
Before tackling this infinite subset, let us see in more details 
how the amplitude behaves for the zero mode part where $\Ket{B} = f\Ket{0}$.
We can get the zero mode part of the amplitude very simply:
\begin{align*}
\mathcal{A}_{0,0}(E) & =\int dx_0 \int ds \sum_{n=1}^\infty (-\tl)^n e^{ix_0(n-E)} q^{n^2/4}.
\end{align*}
In the last expression we got rid of the unphysical divergent $\Bra{0}q^{L_0}\Ket{0}$.
We then fourier transform as explained before and integrate over $x_0$ and $E$
\begin{align*}
\mathcal{A}_{0,0}(\tau) & =\int ds \sum_{n=1}^\infty (-\tl)^n e^{in\tau} q^{n^2/4},\\
\mathcal{A}_{0,0}(\tau) & = 2 \sum_{n=1}^\infty \frac{(-\tl e^{i\tau})^n}{n^2},\\
& = 2\text{Li}_2(-\tl e^{i\tau}),
\end{align*}
where Li$_2(z)$ is the polylogarithm function.
We can now Wick rotate $\tau = -it$ and fourier transform back:
\begin{align*}
\mathcal{A}_{0,0}(E) & = 2 \int dt e^{iEt} \text{Li}_2(-\tl e^t).
\end{align*}
The following equation relating the polylogarithm to the Fermi-Dirac 
distribution is very useful:
\begin{align}\label{polylog}
\text{Li}_{(1+s)}(-e^\mu) & = \frac{-1}{\Gamma(s+1)}
\int_0^\infty \frac{k^sdk}{e^{k-\mu}+1}.
\end{align}
Using this, we can write the amplitude in a way very similar to \cite{Lambert:2003zr}, 
and the integral over time leads to a very similar result
\begin{align}
\mathcal{A}_{0,0}(E) & = -2 \int_0^\infty dk \int dt \frac{k e^{iEt}}{\frac{1}{\tl} e^ke^{-t} +1},\\
\mathcal{A}_{0,0}(E) & = - 2 \int_0^\infty dk ke^{iEk} e^{iE\log(\tl)}
\frac{\pi i}{\sinh{(\pi E)}},\nonumber\\
& = - e^{iE\log(\tl)}\frac{2\pi i}{\sinh{(\pi E)}E^2}.\nonumber
\end{align}
At this stage we have what we expected, this is the same as the tree level modified by a factor
of $1/E^2$ coming from the closed string propagator. One can then generalize to higher levels and we will do so in more details in the following section.  

Two complications arise when one look at this expansion to higher levels.
The first is that one finds generically some hypergeometric functions that are, in general, difficult to integrate over time.  This is only a technical problem as the integrand has 
no divergences other than poles.

The second complication is more important and the main source of worry coming from this amplitude. It is the fact that at level 2 or higher we start getting divergent terms that cannot be integrated of the form 
(\ref{BadIntegral}).  We will now show how these terms can be summed into something finite.

\section{Partial Summation}
At this point, our approximation is clearly unsatisfying. 
Here we will upgrade the approximation by doing a partial summation over all levels.  We will
use the fact that we know the boundary state coefficient of the form (\ref{infinitesub}) to all levels to 
compute the amplitude for all this terms.  So we take the boundary state to be of the form:
\begin{align}
\Ket{B} & = \left(B^{(0;0)} + \sum_{N=1}^{\infty}\frac{1}{\sqrt{N}} B^{(N,N)}\a_{-N}\ta_{-N}\right)\Ket{0},\\
B^{(N;N)} & = \sum_{n=0}^\infty \tg^n -\frac{2}{N}\sum_{n=0}^{N-1}(N-n)\tg^n.\nonumber
\end{align}
To evaluate this we will
need to expand the vertex operator and the boundary state in the $\a$ basis
\begin{align}
e^{-iEX^0} & = e^{-iE(x_0 + i\sqrt{2}\sum_{n >0} \frac{1}{n}(\a_n +\ta_n))} 
 = e^{-iEx_0}\left(1+2E^2\sum_{n>0}^\infty \frac{1}{n^2}\a_n\ta_n + \dots\right),\\
\Bra{N} & = e^{- \sum_{n>0} \frac{1}{n}\a_n\ta_n} 
 = 1 - \sum_{n>0}^\infty \frac{1}{n}\a_n\ta_n+\dots\;,\nonumber
\end{align}
where we have used the Neumann boundary condition ($(\alpha_n + \tilde\alpha_{-n})\KKet{N} = 0$) 
for the vertex operator and
we have also use the fact that $\eta^0_0 = +1$ in the euclidean theory for the Neumann boundary 
state.
Commuting all the $\a$ we get 
\begin{align*}
\Expect{e^{-iEX_0}} & = e^{-iEx_0}\left(\sum_{n=0}^\infty \tg^n q^{n^2/4} \right.\\
&\left.+
\sum_{N=1}^\infty \sqrt{N}\left(\frac{2E^2}{N} -1\right)q^N \left[
\sum_{n=0}^\infty \tg^n q^{n^2/4}
-\frac{2}{N}\sum_{n=0}^{N-1}(N-n)\tg^nq^{n^2/4}\right]\right).
\nonumber
\end{align*}
Then we take the fourier transform and intregate over $x_0$, E and s (in this order)
\begin{align}
\A{}(\tau) & = \A{0,0}(\tau) + \sum_{N=1}^\infty \frac{1}{\sqrt{N}} \left[
\hf \sum_{n=0}^\infty \frac{2n^2 -N}{n^2+4N} z^n -\frac{1}{N}
\sum_{n=0}^{N-1} (N-n)\frac{2n^2 -N}{n^2 +4N} z^n\right],
\end{align}
where $z=-\tl e^{i\tau}$. This includes $\A{0,0}$ calculated previously.  The two sums over 'n' can be
done but the answer is quite messy as one gets many different terms involving hypergeometric function
of the type $_2F_1(1, -2i\sqrt{N}, 1-2i\sqrt{N}, z)$. The $\sqrt{N}$ makes the sum over N intractable
analytically. These terms are not really worrisome but difficult to analyze to the end. 
As we are already considering only a subset of
terms contributing to the amplitude, there is no reasons to even bother with 
them and so we will just drop them.  Keeping only the part of the sum that can be done analytically:
\begin{align}\label{Approx}
\sum_{n=0}^\infty \frac{2n^2 -N}{n^2+4N} z^n  & = \frac{2}{1-z} + \dots,\\
 \sum_{n=0}^{N-1} (N-n)\frac{2n^2 -N}{n^2 +4N} z^n & =
 \frac{-2z(1-z^N)}{(1-z)^2} + \frac{2N}{1-z} + \dots\;.\nonumber
 \end{align}
And, in this approximation the amplitude is 
\begin{align}
\A{}(\tau) & \approx \A{0,0}(\tau) + 4\sum_{N=1}^\infty \frac{1}{\sqrt{N}}\left[
\frac{-1}{1+\tl e^{i\tau}} + \frac{2}{N}\left( \frac{\tl e^{i\tau}}{(1+\tl e^{i\tau})^2} +
\frac{(-\tl e^{i\tau})^{N+1}}{(1-\tl e^{i\tau})^2}\right)\right].
\end{align}
The first term is divergent but this is an artifice of our approximation (\ref{Approx}) as we expect that
there should be other N dependent terms and the sum over N would then give a number. For the moment we simply regulate this sum using the Riemann zeta regulation.
The second term in the sum is the source of all the worries mentioned at the beginning of 
this paper. For level 1, it is finite but for higher levels each term individually would diverge after inverse Wick
rotation.  But here comes the important point: the sum over oscillator levels is convergent before inverse Wick
rotation, yielding a finite result for the amplitude
\begin{align}
\A{\rm{partial}}(t) & \approx Li_2(-\tl e^t) + 4\left[ -\xi(1/2) \frac{1}{1+\tl e^t} \right.\\
& \left.-2\xi(3/2) \frac{-\tl e^t}{(1+\tl e^t)^2} + 2\frac{\tl e^t}{(1+\tl e^t)^2} Li_{3/2}(-\tl e^t)\right].\nonumber
\end{align}
Now we need to take the inverse Fourier transform. Doing so we see that each integral is convergent as we have more power of $e^t$ in the denominator than in the numerator.  Each integral gives a factor 
of $1/\sinh{\pi E}$.  The last integral is a little more involved and we
need to do a numerical integration for the dummy variable.  It is simple to show(for large E) that one gets the following form for the amplitude:
\begin{align}\label{final}
\A{\rm{partial}}(E) & \approx \frac{\pi i e^{-iE\rm{ln}\tl}}{\sinh{\pi E}}w(E),
\end{align}
where $w(E)$ is a polynomial.
The answer is independent of $\lambda$ up to an irrelevant phase as it must be for the 
$\hf$ S-brane.
Now clearly, we cannot take $\A{\rm{partial}}$ too seriously as we calculated it by truncating the boundary state.  By doing so, we have broken the $SU(2)$ symmetry and we probably lost
many other important contributions to the amplitude. 
Nevertheless, we see here that this part of the amplitude 
is finite after \emph{we sum over both momentum and oscillator levels} generalizing the methods and results of Sen to 1-loop. A similar conclusion was obtained previously by Okuyama \cite{Okuyama:2003jk}
for some specific vacuum amplitude using different techniques.\footnote{We are grateful to Kazumi Okuyama for bringing this to our attention.}

This conclusion is expected 
to be true from T-duality since there should be no difference in our treatment
of winding number and oscillator levels.
This gives good hints that the worrisome terms in Sen's boundary state do not lead to inconsistencies. Unfortunately, without a complete expansion of the boundary state into the '$\alpha$' basis, it is impossible to push this calculation further.  In the appendix we explain how one can use the $SU(2)$ symmetry to perform the calculation.
 
\section{Conjecture on the 2-point 1-Loop Amplitude}
 
Finally, let us mention that it is quite unsatisfying that we have still such a limited knowledge of what D-branes decays into.  It is important to realize that the common statement that D-branes decay completely into closed strings radiation is true only if there are no other D-branes around (and small $g_s$).  The first correction to this behavior comes from the 1-loop open string production that we have investigated in this paper.  There will be further corrections of order $g_s$ (see figure 5). 
 \begin{figure}[h]
\centering
\includegraphics[width=7cm]{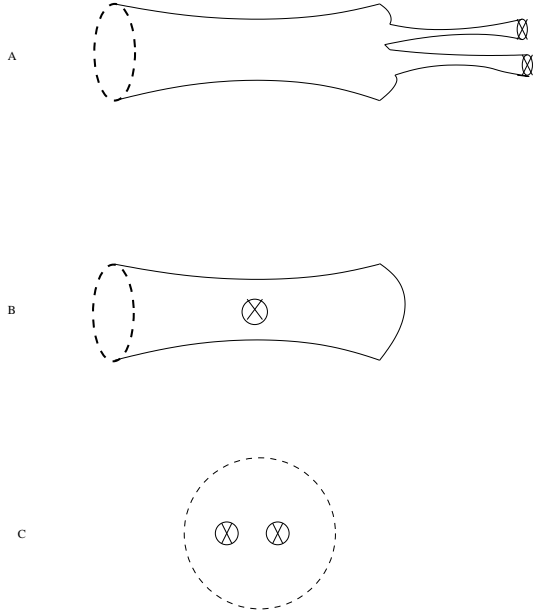}
\caption{Next to next leading order correction to brane decays.}
\end{figure}
Of all these additional diagrams, the two-point function of open strings looks the most interesting. Pair production of open strings might be energetically favorable over coherent production in certain cases.
Here let us 
analyze the basic property of this process.  Firstly, pair production in a time dependent background is 
different then the coherent production of the 1-point 1-loop amplitude and the 1-point tree level calculation that we have discussed in this paper.  Indeed, for the latter two processes the out state is a coherent state and one 
can calculate the total number of particles produced as well as the total energy emitted to be  
\cite{Lambert:2003zr}:
\begin{align}
\frac{\bar N}{V_p} & = \sum_s \frac{1}{2E_s} |\A{s}|^2,\\
\frac{\bar E}{V_p} & = \sum_s \hf |\A{s}|^2.
\end{align}
On the other hand pair production lead to a {\it squeezed state} .
Pair production of particles in a time dependent background has been
studied extensively in the literature (see \cite{Jacobson:2003vx} for
a review on the subject). The out state is:
\begin{align}\label{Squeezed}
  &\Ket{0}^* = 
  \prod_{\bk} \left[\N_\bk \exp\left(-\frac{\alpha_\bk}2
    a^\dagger_\bk a^\dagger_{-\bk}\right)\right]\Ket0,
  &|\N_\bk|^2 = |1-|\alpha_\bk|^2|,
\end{align}
where $a^\dagger_\bk$ is the spacetime creation operator for a bosonic
state of momentum $\bk$.  The two-point amplitude for bosonic
particles (out states) which couple to the time-dependent background
is
\begin{align*}
  \mathcal A_\bk \equiv \Expect{\phi(\bk)\phi(-\bk)}^*
  = \BraKet{\bk,-\bk}{0}^* 
  = \frac{\Bra{0}\sqrt{2E_\bk}a_\bk\sqrt{2E_{-\bk}}a_{-\bk}\Ket{0}^*}
  {\BraKet{0}{0}^*}
  = -(2E_\bk)\alpha_\bk.
\end{align*}
Also, the expectation of {\it out} particle number in the squeezed
out-state created by the {\it in} vacuum is \cite{Birrell:1982ix} (and
\cite{Jacobson:2003vx} for a recent review)
\begin{align*}
  \Expect{N_\bk} &= \frac{|\alpha_\bk|^2}{|1-|\alpha_\bk|^2|},\\
& = \frac1{4E_\bk^2}\frac{|\mathcal A_\bk|^2}
{|1-|\frac{\mathcal A_\bk}{2E_\bk}|^2|}.
\end{align*}
From this we can write the total number of particles emitted as well 
as the total energy emitted
\begin{align}\label{TotalNumber}
& \frac{\bar N}{V_p} = \sum_\bk \frac1{4E_\bk^2}\frac{|\mathcal A_\bk|^2}
{|1-|\frac{\mathcal A_\bk}{2E_\bk}|^2|},
& \frac{\bar E}{V_p} = \sum_\bk  \frac1{4E_\bk}\frac{|\mathcal A_\bk|^2}
{|1-|\frac{\mathcal A_\bk}{2E_\bk}|^2|}.
\end{align}
It is understood in the previous formula that the sum is replaced 
by an integral
for continous degrees of freedom. 
The other important aspect of this process is that the kinematics are completely similar to the closed 
string case. \emph{Pair production of open strings has the same kinematics as coherent production of closed
strings.} 
Indeed, the 2-point amplitude looks roughly like
\begin{align}
\Expect{e^{iEX^0(\omega)}e^{iEX^0(0)}} \approx \Expect{e^{2iEX^0(0)} g(E,\omega)},
\end{align}
where $g(E,\omega)$ is from the contraction and $\Expect{}$ means the cylinder amplitude.  Now, let us make the following ansatz for the 2-point 1-loop amplitude:
\begin{align}
\A{} \approx \frac{1}{\sinh{2\pi E}} w(E),
\end{align}
where $w(E)$ is a polynomial and let us assume that it depends only weakly on the specific state considered. Note that this is exactly the form we found for our partial answer for the 
1-point function with $E \rightarrow 2E$. At large energy, the square of this amplitude behaves like
\begin{align}
\A{}^2 \sim e^{-4\pi E}
\end{align}
The density of state for the bosonic open string is
\begin{align}
D(n) & \sim \frac{1}{\sqrt{2}}n^{-27/4}e^{4\pi\sqrt{n}}\\
E &= \sqrt{\bk_\parallel^2 + n }\sim 
\sqrt{n} + \frac{\bk_\parallel^2}{\sqrt{n}}\nonumber
\end{align}
As in the case of closed string \cite{Lambert:2003zr}, the 
exponentials cancels exactly. This is not surprising. Emitting two 
open strings or one left/right identical closed string lead to the 
same energetic behavior. 

There is still some difference as the phase space is now the world-volume of the brane 
and the final state is a squeezed state as opposed to a coherent state.
Although we can neglect the extra term in the denominator in (\ref{TotalNumber})
we still have an extra factor of E in the denominator compared to 
the closed string
case.  Integrating over the world-volume dimensions of the brane will 
give an extra factor of $n^{1/4}$ per parallel direction, so the phase 
space is the opposite of the one for closed string emission.
Hence, just from phase space considerations, higher 
dimensional branes have their open string production enhanced.

Altogether the results for our ansatz amplitude are 
\begin{align}\label{bosonic}
& \bar N \propto N^2g_s^2 \sum_n n^{(-31 +p)/4} w(\sqrt{n})^2, 
& \bar E \propto N^2g_s^2 \sum_n n^{(-29 + p/4)}w(\sqrt{n})^2.
\end{align}
It is important to note that the for the superstring case, the exponential again cancels but the power law will differ (as does the allowed value for p).  Indeed for the superstring case we get (with 
$D(n) \approx n^{-11/4}e^{4\pi \sqrt{n}}$),

\begin{align}\label{susy}
& \bar N \propto N^2g_s^2 \sum_n n^{(-15 +p)/4} w(\sqrt{n})^2, 
& \bar E \propto N^2g_s^2 \sum_n n^{(-13 + p/4)}w(\sqrt{n})^2.
\end{align}

At this point let us summarize the physical picture that we are getting:
\begin{itemize}
\item We can compare the average number of open strings emitted to what one get from closed strings
emission at tree level \cite{Lambert:2003zr}, 
\begin{align}\label{closed}
& \bar N \propto \sum_n n^{-p/4-1},
& \bar E \propto \sum_n n^{-p/4 +2}.
\end{align}
We see that this is very similar except that the phase space is inverted.

\item The expressions (\ref{closed}) is divergent for $p=0,1,2$ where it is believed that the backreaction becomes 
important and one must regulates the answer.  It was also argued in the litterature that we must always allow for the maximum phase space available (the D0 case). The argument is that the tachyon field should decay in a non-uniform way and that it should be possible to emit closed strings with momemtum  parallel to the brane as well as perpendicular and therefore the phase space allowed for closed string production should correspond to the full nine spatial dimensions.

\item For open strings, it is impossible for them to have momemtum in the perpendicular direction and therefore the phase space allowed, except for the D9 case, is suppressed compare to the closed string case.  This seems to say that open string production would be negligible compare to closed string production because of phase space.  Of course, it is possible that the unknown polynomial $w(\sqrt{n})$ makes the total number and total energy diverge for open strings on a D0-brane as well (to do so we need a strong energy dependence $w(\sqrt{n}) \approx n^{14}$ for the superstring case which appear unlikely). 

\item Finally, the emission of closed strings will be further enhanced by its own 1-loop contribution. This is very similar to the open strings amplitude except that we have two propagators.  Naively, we expect that this extra propagator will suppress the amplitude by an extra factor of $1/E^2$ but more detailed calculation are needed here. Furthermore, there is also a 2-points tree level diagram
that contributes to closed string production at order $g_s$. This diagram is 
not energetically supressed by propagators but neither is it enhanced by the
number of spectators branes. 
\end{itemize}

\section{Conclusion}

In this paper, we have laid down the very first step of a tedious but important calculation. 
We have shown that  divergent terms appear in the amplitude when we expand it in terms of
oscillator levels.  We have also seen that unlike the tree level case, these terms cannot be 
'gauged away'.  Nevertheless, we have obtained partial results
for the amplitude where we considered only a set of powers of q and summed over
all oscillator levels and momenta.  We have shown that by doing so, one can obtain a finite 
amplitude out of pieces that diverge when looked at individually. This is not
suprising as from T-duality, we would expect that the winding number (momenta)
and oscillator levels are interchangeable.  We know that we need to sum over
all momemtum before inverse Wick rotating in the tree level calculation 
and in this paper we have shown that
the same is true (for at least a partial subset) for the oscillator mode in 
the 1-loop calculation. 

Importantly we have also clarified where to perform the inverse Wick rotation. It is crucial to inverse Wick rotate only the final answer $\A{}(\tau)$ as this is the physicially meaningful quantity. Futhermore, the integral over the moduli space 's' and the sum over momentum
and oscillators must be performed in the euclidean theory because all three operations
converge only in that case.  

We have also learned that the known techniques to do 1-loop amplitude are somehow insufficient to
obtain the answer.  Even, if we could succeed in finding the amplitude to the simplest state, what we are
really interested in is the amplitude to any state and the calculation looks almost hopelessly complicated
in that case.  Considering the phenomenological 
importance of this calculation, we need more progress on this issue.  
We also confirmed that Sen's boundary state is really the correct boundary state and these ill-defined 
terms do not cause problem as in a any physical computation only finite results are obtained. 

At this point, considering the form of the partial amplitude it is tempting to conjecture that the 1-point and 2-point open string production have the following form:
\begin{align*}
&\A{1\rm{pt}}(E) \approx \frac{1}{\sinh{\pi E}} w(E), 
& \A{2\rm{pts}}(E)\approx \frac{1}{\sinh{2\pi E}} w^\prime(E),
\end{align*}
where $w(E)$ and $w^\prime(E)$ are \emph{polynomials} and most probably they are functions of the specific state considered as well. Indeed, unlike for the closed string case, there is no reason to expect that the amplitude is the same for all states at a given level.  If this conjecture is true then one can show simply (if $w^\prime(E)$ varies weakly for different states at large energy) that for the 
2-point function the exponential suppression is exactly cancelled by the exponential growth of states.
Then, and only then, can we have a production of open strings that is 
non-zero at high energy. Indeed, in the case where the exponential suppression is less then the  Hagedorn exponential growth, we would have 
negligible open string production, and in the case where it is
more we would have an exponentially divergent amplitude at large energy! Since the latter is unphysical and the former appears counter-intuitive as we expect the spectator branes to be excited, we believe this conjecture to be sensible. From here it then remains to find these unknown polynomials.

Although open strings production is not negligible by itself (it is not zero) 
it will be negligible when compared to closed strings production because of phase space consideration.  Closed strings can go anywhere, in the bulk or the branes, and this enhances their production. There is a slight chance that the polynomial $w^\prime(E)$ is just so such that the open string production does compete even after phase space consideration but this appears unlikely. 

\section{Acknowledgment}
Nicholas Jones has been 
participating at various stage in this work. His deep and insightful comments were
invaluable. We would also like to thank Xingang Chen, Maxim Perlstein, Saswat Sarangi, Sarah Shandera, Ben Shlaer, Ira Wasserman, Piljin Yi and more especially Henry Tye 
for useful comments and discussion. 
This work was supported by National Science Foundation under Grant No. PHY-009831.

\section{Appendix - SU(2) methods}
\subsection{Current Algebra}

In the main body of the text, we have evaluated the one-point function by expanding both the vertex 
operators and the boundary state in the Virasoro basis. As the boundary state is only partially known in 
this basis, only partial answers were obtained.

Another approach is possible that has the important advantage of considering the whole 
boundary state.  We can use SU(2) current algebra or its fermionized version to express the vertex
operators in an SU(2) invariant way and then use the SU(2) basis for the boundary state.  This is possible
as the boundary state only excites open strings with their momenta at an integer value of the SU(2) radius.

Indeed, the amplitude (\ref{theamplitude}) can be written in the following way:
\begin{align*}
\A{Eu}(E) & = \int dx_0 \Expect{e^{-iE X^0}} = 2\pi \sum_n \delta(E-n)\Expect{e^{-iEX^0_{\rm osc}}},\\
\A{Eu}(\tau) & = \sum_n \Expect{e^{-in \tau}e^{-in X^0_{\rm osc}}}, 
\end{align*}
where $\Expect{}$ is the one-point function on the cylinder.

To evaluate $\Expect{e^{-in\tau}}$ we can either write the vertex in terms of SU(2) current operators
($J$) or in terms of fermions.  
For the first case, the following representation of the boundary state seems appropriate
\begin{align}
\KKet{B} &= e^{-\lambda J^+_0}\KKet{N,}\\
\KKet{B} & = \sum_j\sum_{m \geq 0}^j \frac{(j+m)!}{(2m)!(j-m)!}(-\lambda)^{2m}.
\KKet{j,m,m}
\end{align}
The $SU(2)$ generators are defined by:
\begin{align}
J^\pm & =\oint \frac{dz}{2\pi i} e^{\pm 2iX}(z)\\
J^3(z) & = \oint \frac{dz}{2\pi i} i\partial X(z)
\end{align}
As is usual we are interested in the affine lie algebra $SU(2)_1$.
We can express the vertex operator in terms of these operators for different 
$SU(2)$ sectors ($q$) There is an extra factor of 2 from the Neumann boundary condition $X = X_L+X_R
\rightarrow 2X_L$.
\begin{center}
\begin{tabular}{ccc}
$q=1$ & $e^{-iX}(z)$ & $J^-(z)$\\
$q=2$ & $e^{-i2X}(z_1)$ & $z_{12}^{-2} J^-(z_1) J^-(z_2)$\\
$\vdots$ & &\\
$q=n$ & $e^{-inX}(z_1$) & $\Pi_{i,j=1,i<j}^n z_{ij}^{-2} J^-(z_i) J^-(z_j)$
\end{tabular}
\end{center}
For a given sector, only one term in the
exponential will contribute as the number of $J^-$ and $J^+$ must match in order
to have a non-zero amplitude.  So the general amplitude for a given q sector will
be proportinal to $\lambda^q$ (this is the delta function of energy that we were getting
previously). Furthermore, the $J^-$ can be expanded in terms of modes
\begin{align}
J^-(z) = \sum_m J^-_m z^{-m}.
\end{align}
As the amplitude should be independent of $z$, only a finite number of terms would contribute in
the previous sum.   The remaining computation then amounts to  commuting all the SU(2) operators and
acting on the $SU(2)$ Ishibashi state.  This is simply stated but becomes quite messy in practice.

\subsection{Fermionization and Path integral}

Here let us just discuss two alternatives methods that might prove to be the key to making this
calculation tractable.  We do not give any details but merely indicate the general philosophy behind
these two methods.

The $SU(2)_1$ system (or $SU(2)_2$ for superstring) allows for fermionic representations 
(alternatively $SO(2)$ and $SO(3)$).  This approach was used in \cite{Polchinski:1994my} to calculate
the vacuum amplitude and it was further generalized in \cite{Lee:2005ge} to the superstring case.

This approach has the distinct advantage of having the simplest version of the boundary state
but as for the current algebra methods it quickly gets complicated for large momenta and the sum over
'n' appears to be intractable \cite{nick}.  This method is similar in spirit to the previous calculation though
the grassmanian nature of the variable simplifies the algebra.  

Finally, as an alternative to the BCFT methods presented in this paper, one might wonder if one could just use directly path integral techniques to perform the calculation \cite{Larsen:2002wc, nick}. 
This has been shown to be successful at tree-level and together with matrix techniques
one can calculate the one-point function on the disc for various vertex operators.  This is how 
\cite{Constable:2003rc} obtains the coefficients of the boundary state that we have used.
Here the idea is to expand the exponential of the boundary action and to evaluate each correlation
function on the cylinder.  The results are $\Theta$ functions that then need to be integrated,

\begin{align}
& \int dx^0 \Expect{e^{-iEX^0} e^{-\tl \int d\tau e^{ix^0} e^{iX^\prime(\tau)} } },\\
& = \int dx^0 \int ds \sum_{n=0}^\infty e^{-iEx^0}\frac{(-\tl e^{ix^0})^n}{n!}\int d\tau_1 \cdots \int d\tau_n 
\Expect{e^{-iEX^0} e^{-iX^\prime(\tau_1)} \cdots e^{-iX^\prime(\tau_n)}}.\nonumber
\end{align}
 \begin{figure}[h]
\centering
\psfrag{0}{$t_0$}
\psfrag{1}{$t_1$}
\psfrag{2}{$t_2$}
\psfrag{n}{$t_n$}
\includegraphics[width=7cm]{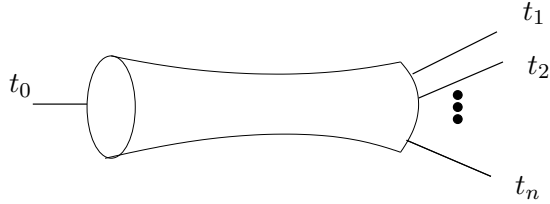}
\caption{One point function where the brane anti-brane boundary is represented by n tachyons.  The full amplitude is obtained by summing over all such diagrams with different n.}
\end{figure}
  This amplitude is simple to evaluate in general but the integrals over time are quite difficult
 \begin{align*}
 \A{}(\tau) & = \int ds \sum_{n=0}^\infty \frac{(-\tl e^{i\tau})^n}{n!} (\rm{z.m.}) 
 \int d\tau_0 \cdots d\tau_n \prod_{i<j}^n \left|\TF{1}{i\tau_{ij}}{is}\right|^2
 \prod_{i=1}^n\left|\TF{2}{i\tau_{0i}}{is}\right|^{-2n}.
 \end{align*}
 Here (z.m) is the zero mode part. All these methods are consistent in that they do not have any obvious simplifications and they quickly lead
 to quite complicated expressions.  Hopefully, further research might find either an alternative method
 or some relations that allow us to complete this important calculation.

\bibliographystyle{JHEP}
\providecommand{\href}[2]{#2}\begingroup\raggedright\endgroup

\end{document}